\documentclass[12pt]{article}

\usepackage{textcomp}
\usepackage{chetnew}
\usepackage{amssymb,amsmath,amsfonts,manfnt}
\usepackage[utf8]{inputenc}

\renewcommand{\d}[1]{\ensuremath{\operatorname{d}\!{#1}}}

\def\one{{\,\hbox{1\kern-.8mm l}}}

\newcommand{\SO}{\mathrm {SO}}
\newcommand{\SU}{\mathrm {SU}}
\newcommand{\U}{\mathrm U}
\newcommand{\Tr}{\mathrm{Tr}}

\newcommand{\CN}{\mathcal{N}}

\usepackage{IEEEtrantools}
\allowdisplaybreaks

\def\makeatletter{\catcode`\@=11}
\makeatletter
\def\mathbox#1{\hbox{$\m@th#1$}}%
\def\math@ccstyles#1#2#3#4#5#6#7{{\leavevmode
      \setbox0\mathbox{#6#7}%
      \setbox2\mathbox{#4#5}%
      \dimen@ #3%
      \baselineskip\z@\lineskiplimit#1\lineskip\z@
      \vbox{\ialign{##\crcr
             \hfil \kern #2\box2 \hfil\crcr
             \noalign{\kern\dimen@}%
             \hfil\box0\hfil\crcr}}}}
\def\mathaccstyles{\math@ccstyles\maxdimen}
\def\maththroughstyles{\math@ccstyles{-\maxdimen}}
\def\unity%
 {\maththroughstyles{.45\ht0}\z@\displaystyle {\mathchar"006C}\displaystyle 1}

\preprint{QMUL-PH-16-02}

\title{The NS limit of the 5D Superconformal Index}

\author{Constantinos~Papageorgakis,$^{1\diamondsuit}$ Alessandro~Pini,$^{2\clubsuit}$ and Diego~Rodríguez-Gómez$^{2\heartsuit}$}

\affiliation{\\$^1$ CRST and School of Physics and Astronomy\\ Queen
  Mary University of London, E1 4NS, UK\\ $ $ \\$^2$ Department of Physics,
  Universidad de Oviedo\\Avda. Calvo Sotelo 18, 33007, Oviedo, Spain

\emails{$^{\diamondsuit}$c.papageorgakis@qmul.ac.uk,$^{\clubsuit}$pinialessandro@uniovi.es,$^{\heartsuit}$d.rodriguez.gomez@uniovi.es}
}

\abstract{We consider the Nekrasov-Shatashvili (NS) limit of the
  five-dimensional superconformal index and propose a novel
  prescription for selecting the finite contributions.  Applying the
  latter to various examples of $\U(1)$ theories, we find that the 5D
  NS index can be reproduced using recent techniques of Córdova and
  Shao, who related the 4D Schur index to the BPS spectrum of the theory
  on the Coulomb branch. In this picture, the 5D instanton solitons
  are interpreted as additional flavour nodes in an associated 5D BPS
  quiver.}

\date{\today}

\begin{document}

\maketitle

\toc

\section{Introduction and Summary}

The superconformal index has proved to be an important tool in the
study of superconformal field theories (SCFTs) in diverse dimensions
\cite{Romelsberger:2005eg,Kinney:2005ej,Bhattacharya:2008zy}. In some
cases interesting limits of the index have been devised, which isolate
contributions from particular subsets of operators and provide
information about its different phases, see {\it e.g.}
\cite{Gadde:2011uv}. Limits of the index also help in identifying
algebraic structures hidden within special subsectors of the theory, a
fact which has been put to remarkable effect in four and six
dimensions \cite{Beem:2013sza,Beem:2014kka}.

In a closely related direction, recent work \cite{Cordova:2015nma}
established a connection between the so-called Schur limit of the 4D
index on the one hand,\footnote{Recent exact results on the 4D Schur
  index include
  \cite{Bourdier:2015wda,Bourdier:2015sga,Drukker:2015spa}.} and a
certain algebraic quantity associated with the BPS spectrum of
particles on the Coulomb branch on the other---the trace of the
Kontsevich--Soibelman (KS) operator---for a convincing number of 4D
$\mathcal N=2$ SCFTs;\footnote{For an alternative calculation of the
  Schur index for Argyres--Douglas theories see
  \cite{Buican:2015ina}.} see also \cite{Cecotti:2015lab} for
generalisations. In this fashion one demonstrates that, for specific
BPS subsectors, the operator spectrum of an SCFT is directly related
to the particle spectrum of the same theory in a phase where the
conformal symmetry has been broken.

In this note, we would like to import some of these results to
five-dimensional SCFTs
\cite{Seiberg:1996bd,Morrison:1996xf,Douglas:1996xp,Intriligator:1997pq,
  Aharony:1997bh}. Our first objective will be to define a limit of
the 5D superconformal index by turning off one of the two
$\Omega$-deformation parameters;\footnote{Since the precise operator
  spectrum of the interacting 5D UV theories is unknown, one usually
  works with the realisation of the index as a supersymmetric
  partition function on $S^4 \times S^1$ with twisted boundary
  conditions for the various fields.} this is the limit first
considered by Nekrasov and Shatashvili (NS) in a four-dimensional
context \cite{Nekrasov:2009rc}. Its naive implementation leads to a
singular index, which calls for a prescription on how to extract the
finite parts. This problem can in principle be addressed in a way
similar to the original NS limit of \cite{Nekrasov:2009rc}. However,
the direct 5D extension of that recipe leads to a function whose
fugacity expansion does not necessarily involve integer
coefficients. In turn, we propose a different 5D regularisation which
results in a fugacity expansion with integer coefficients for
arbitrary gauge groups. In the abelian case, our regularisation
clearly isolates contributions from states localised on a
four-dimensional subspace of the euclideanised spacetime. Moreover, it
reproduces, at least for the perturbative sector, the large-orbifold
limit of the gauge theory index of \cite{Mekareeya:2013ija}. The
latter effectively reduces the space down to a 4D geometry---of the
form $M_3 \times S^1$---where the contributions of vector and
hypermultiplets become identical to the 4D Schur index and may hint
towards an interesting connection with
\cite{Beem:2013sza,Beem:2014kka}. Although our limit does not lead to
a counting of states preserving a larger fraction of
supersymmetry,\footnote{Interesting limits of the 4D index with
  additional superymmetry were originally considered in
  \cite{Gadde:2011uv}.} it does lead to a factorisation of the index
into a ``holomorphic'' and ``antiholomorphic'' part for general 5D
SCFTs. This factorisation is reminiscent of the work of Iqbal and Vafa
\cite{Iqbal:2012xm}, where it also appeared as the starting point for
connecting the 5D BPS-particle degeneracy\footnote{Note that in 5D
  there also exist BPS strings.} to the index, using the topological
string.

With this last point in mind, our second objective will be to relate
the NS limit of the 5D index to the work of
\cite{Cordova:2015nma}. For a number of abelian examples we will show
that the NS index can be reproduced by the trace of the KS operator
for a ``5D BPS quiver''. This quiver can be constructed
straightforwardly by assigning a node for each ``partonic BPS
state''.\footnote{By this we mean states with the lowest possible
  charges, i.e. ones that cannot be written as bound states of any
  other states.} This involves a node corresponding to the
instanton-soliton parton of the 5D theory, as well as a node for each
of the possible $N_f$ hypermultiplets of the theory. The construction
and study of the 5D BPS quiver for nonabelian theories, and their
possible connection to the NS index, is a question that we will leave
open for future investigation. However, our abelian results can
already be thought of as a check of the proposal of
\cite{Iqbal:2012xm}, for a particular subsector of five-dimensional
theories.

The rest of this article is organised as follows: In Sec.~\ref{NSNS}
we will present the details of the NS limit for the 5D index, after
briefly reviewing some background material necessary for our
discussion. Then in Sec.~\ref{KSKS} we will introduce the algebraic
tools of \cite{Cordova:2015nma} and use them to recover our index for
$\U(1)$ theories with different matter content and values of the
Chern-Simons coefficient. We will also discuss some directions for
generalising these results to nonabelian gauge groups.

\section{The Nekrasov-Shatashvili limit of the 5D index}\label{NSNS}

\subsection{Generalities}

The superconformal index in five dimensions was first defined in
\cite{Bhattacharya:2008zy} and computed using supersymmetric
localisation \cite{Pestun:2007rz} for a variety of $\mathcal N=1$
theories in \cite{Kim:2012gu}. Recall that using a Verma module
construction, one can obtain all irreducible representations of the 5D
superconformal algebra (SCA) $F(4)$ from irreducible representations
of the maximal compact subalgebra
$\mathfrak{so}(2)_E\oplus \mathfrak{so}(5)\oplus \mathfrak{su}(2)_R$.
The latter are labelled by strings of quantum numbers denoting the
highest weight state $\{\epsilon_0,\,R,\,h_1,\,h_2\}$, where
$h_1,\,h_2$ are the Cartan generators of
$\mathfrak{so}(5)$,\footnote{These are related to the
  $\Omega$-deformation parameters $\epsilon_1, \epsilon_2$ in a simple
  way.} while $\epsilon_0$ is the scaling dimension measured by the
charge under $\mathfrak{so}(2)_E$. Finally, the $\mathfrak{su}(2)_R$
Cartan generator is denoted by $R$.\footnote{As is common in the
  literature, we will use the same symbols for the Cartan generators
  and the corresponding charges, depending on the context.}

In the radial quantisation of the theory, where $S=Q^\dagger$, and for
a particular choice of supercharge,\footnote{We follow the conventions
  and choices of \cite{Kim:2012gu}.} one can define
\begin{equation}
\delta :=\{Q,\,S\}=\epsilon_0-h_1 - h_2-3\,R\;,
\end{equation}
which is a positive-definite quantity. The index is a partition
function counting operators transforming in irreducible
representations of the subalgebra of the SCA that (anti)commute with
the above $Q,S$ (these are $\frac{1}{8}$-BPS) and hence also
$\delta$---or equivalently, irreps of the commutant of
$(Q,S,\delta)$ of the 5D SCA. It is straightforward to see that
$h_1+R$ and $h_2 + R$ commute with the above choice of $\delta$ and as
a result the most general, or ``refined'', index with respect to the
supercharge $Q$ is given by
\cite{Gaiotto:2015una,Bhattacharya:2008zy,Kim:2012gu}
\begin{equation}\label{5dindex}
I={\rm Tr}_{\mathcal H_{\delta=0}}\,(-1)^F\,\,p^{h_1 +
  R}\,q^{h_2 + R}\,\prod_a w_a^{\mathfrak{Q}_a} \,\mathfrak q^k\;.
\end{equation}
Here the trace is taken over the Hilbert space of $\delta = 0$
operators, $F = 2 h_1$ is the fermion number operator, $p$, $q$ are
fugacities keeping track of the elements of the commutant and the
$w_a$ are additional fugacities for commuting charges
$\mathfrak{Q_a}$, corresponding to possible global/gauge
symmetries. One such commuting charge corresponds to a topological
$\U(1)$ symmetry which is always present in the examples we are
interested in: 5D gauge theories possess a conserved current,
$ * J = \frac{1}{8 \pi^2} \rm{tr}(F\wedge F)$, and their spectrum
contains instanton solitons, charged under the associated symmetry.
This global symmetry plays an important role in five dimensions, where
SCFTs with very interesting properties exist \cite{Seiberg:1996bd}: in
many cases it can combine with and enhance other symmetries (flavour,
Lorentz); see {\it e.g.}
\cite{Kim:2012gu,Lambert:2014jna,Tachikawa:2015mha}. Indeed, one can
also include a fugacity $\mathfrak q$ in the index \eqref{5dindex},
which keeps track of the instanton charge $k$, where $|\mathfrak q|=1$.

Via the state-operator map, the 5D index can alternatively be
evaluated by a Euclidean path integral on $S^4 \times S^1$ with
twisted boundary conditions for the various fields according to their
charges \cite{Bhattacharya:2008zy,Kim:2012gu,Terashima:2012ra}. The
index then counts $\frac{1}{8}$-BPS states for the theory on the
sphere. This functional integral can be evaluated in the IR
theory\footnote{For a generic SCFT on $\mathbb{R}\times S^4$ it is
  possible to turn on supersymmetrically a position-dependent YM
  coupling, interpolating between the SCFT and the IR gauge theory
  \cite{Pini:2015xha,Bergman:2016avc}.}  using localisation
\cite{Kim:2012gu} and the answer reduces to a gauge-group integral
over the product of perturbative and nonperturbative contributions,
schematically
\begin{equation}\label{IRdef}
I=\int [\d U]Z^{S^4}_{\rm pert}Z_{\rm nonpert} \;,
\end{equation}
with $[\d U]$ the unit-normalised Haar measure. The nonperturbative
factor can be written as
\begin{equation}
Z_{\rm nonpert}=|Z_{\rm Nek}|^2 \;,
\end{equation}
where $Z_{\rm Nek}$ is the Nekrasov instanton partition function
\cite{Nekrasov:2003rj,Nekrasov:2002qd}. The perturbative
contribution is a modular quantity built out of the weak-coupling
multiplets. The vectormultiplet and hypermultiplet contributions are
given by
\begin{equation}\label{PEf}
 I_{V,\,H}={\rm PE}[f_{V,\,H}]\ ,
\end{equation}
where ${\rm PE}$ refers to the plethystic exponential. The so-called
single-letter indices appearing above in turn read\footnote{For
  definiteness, we will assume that the hypermultiplet is in the
  fundamental of the gauge group.}
\begin{equation}
\label{pert}
f_V=-\frac{p + q}{(1-p)(1-q)}\,\chi_{\rm Adj}\;,\qquad
f_H=\frac{\sqrt{p q}}{(1-p)(1-q)}(\chi_{\Box}+\chi_{ \bar \Box}) \;,
\end{equation}
with $\chi_{\mathcal R}$ denoting the character of a given
representation $\mathcal R$.

\subsection{The NS index}

Having set the stage, we would like to investigate whether there exist
limits of the index \eqref{5dindex} which only receive contributions
from certain sectors of the theory, as {\it e.g.} is the case in 4D
\cite{Gadde:2011uv}. Note that, as opposed to other dimensions, the 5D
index only depends on two fugacities. Moreover, these correspond to
Cartans of $\SU(2)$ symmetries, a fact which underlies the
$(p,\,q)\leftrightarrow (q^{-1},\,p^{-1})$ and
$(p,\,q)\leftrightarrow (q,\,p)$ invariance of the index;
\textit{c.f.}  \eqref{pert}. Thus, it is hard to imagine nontrivial
regular limits as in \cite{Gadde:2011uv}. Yet, this does not exclude
interesting singular limits. In particular, following
\cite{Nekrasov:2009rc}, we will focus on the Nekrasov-Shatashvili (NS)
limit of the index. Generically, the NS limit involves sending one of
the two $\Omega$-deformation parameters to zero, $\epsilon_1 \to 0$,
while keeping the other one, $\epsilon_2$, fixed. These parameters are
chemical potentials for rotations in two real planes,
$\SO(2)_{\epsilon_1} \times \SO(2)_{\epsilon_2} \subset \SO(5)$, and
related to our choice of fugacities through $p = e^{-\epsilon_1}$ and
$q = e^{- \epsilon_2}$. Hence, one can naively implement the NS limit
directly at the level of the index, by considering
\begin{align}\label{NSlimitnaive}
p\to 1\quad\textrm{and}\quad q\to \textrm{fixed}\;.
\end{align}
Although this definition is natural, it leads to divergences as can be
immediately seen by applying it to the perturbative contributions
\eqref{pert}. We therefore need to put forward a modified definition
for taking the NS limit of the 5D index, which leads to finite
contributions.

Towards that end, we follow \cite{Gaiotto:2015una} and rewrite the
index of the full theory on $S^4 \times S^1$ in terms of two
``hemisphere indices'' on $D^4 \times S^1 $ with Dirichlet boundary
conditions, where $D^4\subset S^4$ is half the sphere. The hemisphere
index is in turn defined by
\begin{align}
  \label{eq:1}
   II = Z^{D^4}_{\rm pert}Z_{\rm Nek}\;.
\end{align}
For the example of a single vectormultiplet and a
hypermultiplet in the fundamental representation, the perturbative
piece reads
\begin{align}
  \label{eq:3}
Z^{D^4}_{\rm pert} = {\rm PE}\Big[-\frac{p  q }{(1-p)(1-q)}\chi_{\rm Adj} +
  \frac{\sqrt{p q}}{(1-p)(1-q)}\chi_{\Box} \Big] \;,
\end{align}
where the gauge symmetry of the full index on $S^4 \times S^1$ is to
be understood as a global boundary symmetry.

The full index is then computed by combining two such contributions
and gauging the appropriate diagonal subgroup of said global
symmetries to obtain
\begin{align}
  \label{eq:4}
 I = (I^{4D}_V)^r \int [\d U] \; II \; \overline{II}\;,
\end{align}
where the overline implies that one inverts all gauge/flavour
fugacities. The term
\begin{align}
  \label{eq:5}
  I_V^{\mathrm{4D}} = {\rm PE}\Big[-\frac{p}{1-p}-\frac{q}{1-q}\Big]
\end{align}
 is a purely four-dimensional $\CN = 1$ vectormultiplet contribution
 coming from the boundary and $r = {\rm rank}(G)$ is the gauge group rank.

We are now in the position to define the NS index as follows:
\begin{align}
  \label{NSlimit}
   \mathrm{NS~index:}\qquad II^{\mathrm{NS}}(z_i,\mathfrak q;q) := \mathrm{PE} \Big
  [  \lim_{p\to 1} (1-p)  \;\mathrm{PE}^{-1} [II(z_i,\mathfrak q;p,q)]
  \Big ]\;,
\end{align}
such that 
\begin{align}
  \label{eq:6}
 I^{\mathrm{NS}}(\mathfrak q;q) := \int [\d U] \; II^{\mathrm{NS}}(z_i,\mathfrak q;q) \; \overline{II^{\mathrm{NS}} (z_i,\mathfrak q;q)}\;.
\end{align}
Note that we have stripped off the (divergent in this limit) factors
of $I^{\rm 4D}_V$. We will come back to this below.

We stress that this definition of the NS limit is different from other
versions where the $\rm{PE}$ in \eqref{NSlimit} is traded for a
standard exponential and results in a function whose fugacity
expansion does not necessarily involve integer coefficients; see
\cite{Nekrasov:2012xe,Nekrasov:2013xda,Nekrasov:2009rc}. On the other
hand, Eq.~\eqref{eq:6} does admit an expansion with integer
coefficients, due to the use of the ${\rm PE}$.

In the above the $z_i$, $i=1,\ldots,r$, are gauge/global symmetry
fugacities and the plethystic logarithm, $\mathrm{PE^{-1}}$, is the
inverse of the plethystic exponential, defined as
\begin{align}
  \label{eq:21}
  \mathrm{PE^{-1}}[g(t)] := \sum_{n = 1}^\infty \frac{\mu(n)}{n} \log[g(t^n)]\;,
\end{align}
with $\mu(n)$ the M\"obius function. This factorisation of the
superconformal index in the NS limit is reminiscent of the discussion
in \cite{Iqbal:2012xm}, where the full index was calculated using the
refined topological vertex formalism and related to the counting of
BPS states on the Coulomb branch of the theory. We will see in the
next section that the relationship to ``5D BPS quivers'' can be quantified
for $G = \U(1)$ through the formalism of \cite{Cordova:2015nma}.

\subsubsection*{Perturbative NS limit}

Since our prescription for the NS limit \eqref{NSlimit} factorises
over the perturbative and nonperturbative contributions, let us first look at the former. From \eqref{eq:3}
it is straightforward to deduce that 
\begin{align}
  \label{eq:2}
Z^{D^4,\mathrm{NS}}_{\rm pert} = {\rm PE}\Big[-\frac{q }{(1-q)}\chi_{\rm Adj} +
  \frac{\sqrt{q }}{(1-q)}\chi_{\Box}\Big] \;
\end{align}
and consequently if we only focus on the perturbative sector
\begin{align}
  \label{eq:7}
\int [\d U]Z^{D^4,\mathrm{NS}}_{\rm pert}\overline{Z^{D^4,\mathrm{NS}}_{\rm pert}} =
  \int [\d U]{\rm PE}\Big[-\frac{2q }{(1-q)}\chi_{\rm Adj} +
  \frac{\sqrt{q }}{(1-q)}(\chi_{\Box} + \chi_{\bar \Box}) \Big] \;.
\end{align}
This is tantamount to projecting out states with a nontrivial $x_{++}$
dependence, as can be seen by taking the NS limit directly on the full
5D single-letter indices.

This requires an equivalent prescription for which it is convenient to
introduce fugacities $x=\sqrt{p\,q}$, $y=\sqrt{q/p}$. Note that, after
performing this substitution in equation (\ref{5dindex}), the
exponents of the $x$ and $y$ fugacities are given respectively by
$ h_1 + h_2 + 2R = 2j_1+2R $ and $-h_1 + h_2=-2j_2$. In terms of
these, the NS index for the hypermultiplet can be implemented by
taking $y\rightarrow x$. More precisely
\begin{align}
\label{eq:NSpert}
f^{\mathrm{NS}}_{H}=\lim_{\epsilon_1\rightarrow 0} \epsilon_1 \,f_{H}(x,\,x(1+\epsilon_1))\; .
\end{align}
In this fashion the NS index picks out the coefficient of the
$\frac{1}{\epsilon_1}$ pole in the naive $\epsilon_1\to 0$ limit of
$f_H$. Recall that for the free hypermultiplet the single particle
index $f_H$ can be understood in terms of letter counting using the
state-operator map \cite{Kim:2012gu}.
\begin{table}
\begin{center}
\begin{tabular}{c | c | c | c }
& $\epsilon_0$ &  $(j_1,\,j_2)$ & $R$ \\ \hline
$q$ & $\frac{3}{2}$ & $(0,\,0)$ & $\pm\frac{1}{2}$ \\ 
$\psi$ & $2$ & $(\pm\frac{1}{2},\,0)\oplus(0,\,\pm\frac{1}{2})$ &$ 0$\\
$\partial$ & 1 & $(\pm\frac{1}{2},\,\pm\frac{1}{2})\oplus (0,\,0)$ & 0
\end{tabular}
\end{center}
\caption{The letters in the hypermultiplet and their respective charges.}\label{Tab:1}
\end{table}
Using Table~\ref{Tab:1}, one immediately sees that $f_H$ contains
operators made out of letters of the form
$\partial_{+\pm}^m \mathcal O$; here $\mathcal O$ is a scalar or
fermionic component of the hypermultiplet and the derivatives are
responsible for the factor $(1-p)\,(1-q)=(1-x\,y)(1-\frac{x}{y})$
appearing in the denominator of \eqref{pert}. In the limit
$\epsilon_1\rightarrow 0$, one such derivative becomes of zero weight.
This results in a divergence in the limit $y\rightarrow x$,
originating from an unrefinement in the index which now counts letters
containing arbitrary powers of $\partial_{++}$ with the same weight
(zero). Defining the NS index through selecting the pole in
\eqref{eq:NSpert}, is tantamount to only accounting for the
contribution with no derivatives.

Somewhat surprisingly, the vector multiplet piece can also be given an IR-operator interpretation. In such a scenario, one can understand the single-letter vector multiplet contribution as arising from components of the gaugino plus a tower of infinitely many derivatives. In the limit $\epsilon_1\rightarrow 0$, not only the weight of a derivative but also one of the components of the gaugino become zero. These translate into singularities of the index and our prescription amounts to regularising them by discarding zero-weight letters.

Hence, at the level of implementation, the following single-letter
functions can be used for the perturbative contributions in the NS
limit:
\begin{align}
  \label{eq:8}
f^{\mathrm{NS}}_V=-\frac{2q}{(1-q)}\,\chi_{\rm Adj}\;,\qquad
f^{\mathrm{NS}}_H=\frac{\sqrt{q }}{(1-q)}(\chi_{\Box}+\chi_{ \bar \Box}) \;.
\end{align}
We highlight that these single-letter terms are precisely the vector
and hypermultiplet single-letter index contributions for the
perturbative sector of $\CN =2$ four-dimensional theories in the Schur
limit \cite{Gadde:2011uv,
  Bourdier:2015wda,Bourdier:2015sga,Drukker:2015spa}, which may hint
at a connection with the results of
\cite{Beem:2013sza,Beem:2014kka}. It is also interesting to observe
that the large-orbifold limit of \cite{Mekareeya:2013ija} also led to
perturbative contributions identical to those of the 4D Schur
index.\footnote{Recall that \cite{Mekareeya:2013ija} considered the 5D
  theory on $S^4/\mathbb{Z}_n\times S^1$ in the large-$n$ limit.  This
  effectively dimensionally reduced the space down to a (singular) 4D
  geometry.}

All in all, in the perturbative sector our NS limit discards states
with a dependence on the $x_{++}$ direction on $D^4$, along with the
boundary $\mathcal N = 1$ vectormultiplet contributions
$I_V^{4D}$. This is equivalent to using the single-letter expressions
\eqref{eq:8} directly into \eqref{PEf}. We will next see that this
interpretation extends to the nonperturbative sector for abelian
theories.

\subsubsection*{Nonperturbative NS limit}

The result of the prescription \eqref{NSlimit} on the nonperturbative
piece is somewhat more involved. This is due to the fact that, with
the exception of the abelian case, the Nekrasov partition function
cannot be written as a PE of single-letter contributions but is
evaluated as an expansion in powers of the instanton fugacity
$\mathfrak q$
\begin{align}
  \label{eq:9}
 Z_{\rm Nek} = \sum_{k=0}^\infty \mathfrak q^k Z_{\rm
  Nek}^{(k)}\qquad\mathrm{with}\qquad Z_{\rm Nek}^{(0)} = 1\;.
\end{align}
We will henceforth assume that the NS limit commutes with the
instanton expansion and then use this along with \eqref{NSlimit} to
get
\begin{align}
  \label{eq:10}
   & Z_{\rm Nek}^{\mathrm{NS}}(z_i,\mathfrak q ;q) = \mathrm{PE} \Big
     [  \lim_{p\to 1} (1-p)  \;\mathrm{PE}^{-1} [  \sum_{k=0}^\infty \mathfrak q^k Z_{\rm Nek}^{(k)}(z_i;p,q)]
  \Big ]\cr
&= \mathrm{PE} \Big [\lim_{p\to 1}  
 (1-p) \;\mathrm{PE}^{-1} [1 + \mathfrak q
     Z_{\mathrm{Nek}}^{(1)}(z_i;p,q) + \mathfrak q^2 Z_{\mathrm{Nek}}^{(2)}(z_i;p,q) +
  O(\mathfrak q^3)]  \Big ]\cr
 & =\mathrm{PE} \Big[
 \lim_{p\to 1} (1-p) \Big(\mathfrak q
  Z_{\mathrm{Nek}}^{(1)}(z_i;p,q) + \cr
&\qquad \qquad\qquad\qquad + \mathfrak q^2 \Big(  Z_{\mathrm{Nek}}^{(2)}(z_i;p,q) -\frac{1}{2}
   Z_{\mathrm{Nek}}^{(1)}(z_i;p,q)^2 - \frac{1}{2}
  Z_{\mathrm{Nek}}^{(1)}(z_i^2;p^2,q^2) + 
  O(\mathfrak q^3) \Big)\Big] \cr
& = 1 + \mathfrak q \lim_{p\to 1}  (1- p)
  Z_{\mathrm{Nek}}^{(1)}(z_i;p,q) + \cr
&\qquad \qquad\qquad  + \mathfrak q^2 \lim_{p\to 1} \Big( (1- p)( Z_{\mathrm{Nek}}^{(2)}(z_i;p,q)  -\frac{1}{2}
   Z_{\mathrm{Nek}}^{(1)}(z_i;p,q)^2 - \frac{1}{2}
   Z_{\mathrm{Nek}}^{(1)}(z_i^2;p^2,q^2))\cr
&\qquad\qquad \qquad \qquad  + (1- p)^2 \frac{ 
  Z_{\mathrm{Nek}}^{(1)}(z_i;p,q)^2}{2} +  (1- p^2) \frac{ 
  Z_{\mathrm{Nek}}^{(1)}(z_i^2;p^2,q^2)}{2} \Big) +  O(\mathfrak q^3)\cr
& =: \sum_{k=0}^\infty \mathfrak q^k Z_{\rm Nek}^{\mathrm{NS},(k)}(z_i;q)\;. 
\end{align}
This proposal is obviously applicable to the case of $G = \U(1)$,
where as we will see shortly the instanton expansion can be explicitly
resummed into a PE. E.g. for a pure U(1) theory one has
\begin{equation}
Z_{\rm nonpert}={\rm PE}\Big[\frac{\sqrt{pq}}{(1-p)(1-q)}(\mathfrak q+\mathfrak q^{-1})\Big]\;.
\end{equation}
In that context, the NS limit once again
explicitly counts states which do not have any dependence on the
$x_{++}$ direction.\footnote{One can also ascribe an IR-operator
  interpetation to the abelian instanton partition function, as the PE
  of single-letter contributions from instanton operators
  \cite{Lambert:2014jna,Bergman:2016avc}} However, the definition
\eqref{eq:10} also makes sense for the case of nonabelian gauge
groups, where $Z_{\rm Nek}^{(k)}$ can be expanded in $q$ to yield
terms with integer coefficients, as expected for an index. We have
explicitly checked this to sufficiently high order for $G = \SU(2)$.

As raised above, we should emphasise that a version of the NS limit
for the Nekrasov partition function has already been considered in
\cite{Nekrasov:2012xe,Nekrasov:2013xda}, along the lines of
\cite{Nekrasov:2009rc}. This is a different limit from the one
discussed here, insofar as it involves replacing plethystic
exponentials with exponentials and plethystic logarithms with
logarithms. Our motivation for \eqref{NSlimit} stems from requiring
finite coefficients in the fugacity expansion and mirroring the
definition of the 4D limits of \cite{Gadde:2011uv}, which act directly
on the single-letter indices.

\section{Kontsevich--Soibelman operators and BPS Quivers}\label{KSKS}

Having provided our definition for the NS index, one can establish a
connection with \cite{Cordova:2015nma}. In that reference---see also
\cite{Cecotti:2015lab}---it was conjectured that the 4D Schur index of
a rank-$r$ theory can be recovered in terms of quantities associated
with the BPS quiver of the theory \cite{Cecotti:2010fi} through
\begin{equation}\label{KSoperator}
I_{\mathrm{KS}}=(q)^{2r}_{\infty}\,{\rm Tr}[\mathcal{O}]\;,
\end{equation}
where the Pochhammer symbol is defined as 
\begin{equation}
(q)_0=1,\qquad (q)_{n}=\prod_{k=1}^{n}(1-q^k)\;.
\end{equation}
Here the quantity $\mathcal{O}$ is the Kontsevich-Soibelman (KS)
operator associated with the BPS quiver of the four-dimensional gauge
theory. Such a theory contains a set of BPS particles on the Coulomb
branch labelled by a vector $\gamma$ in the charge lattice
$\Gamma$. Then, for each $\gamma$ one introduces a formal variable
$X_{\gamma}$ obeying a quantum torus algebra
\begin{equation}
X_{\gamma}X_{\gamma'}=q^{\frac{\langle \gamma,\gamma'\rangle}{2}}X_{\gamma+\gamma'}=q^{\langle \gamma,\gamma'\rangle}X_{\gamma'}X_{\gamma}\;,
\end{equation}
where $\langle \cdot,\cdot\rangle$ is the (integer) Dirac pairing of charges in
the lattice $\Gamma$, which can be read off from the BPS quiver. In
terms of these $X_{\gamma}$, the KS operator can be explicitly written as
\begin{equation}
\mathcal{O}=\prod_{\gamma} E_q(X_{\gamma})\;,
\end{equation}
where $E_q$ is the $q$-exponential function
\begin{equation}\label{Eq}
E_q(z)=\prod_{i=0}^{\infty}(1+q^{i+\frac{1}{2}}z)^{-1}=\sum_{n=0}^{\infty}\frac{(-q^{\frac{1}{2}}z)^n}{(q)_n}\;.
\end{equation}
For a theory without flavour, the trace of the quantum torus algebra
is defined by its action on the formal variables $X_\gamma$
\begin{align}
  \label{eq:15}
  \Tr [X_\gamma] = \left\{\begin{array}{c c} 1  & \gamma = 0\\0 &                                                                   \mathrm{otherwise}\end{array}\right.
\end{align}
and extending linearly. For theories with flavour, there exist flavour
charge vectors $\gamma_f$, which have zero Dirac pairing with all other
$\gamma'\in \Gamma$, $\langle \gamma_f, \gamma'\rangle = 0$.  Morevover,
the definition of the trace needs to be modified to
\begin{align}
  \label{eq:16}
  \Tr [X_\gamma] = \left\{\begin{array}{c c} \prod_i
                            \Tr[X_{\gamma_{f_i}}]^{f_i(\gamma)}  &
                                                                   \langle\gamma,\gamma'\rangle
                                                                   =
                                                                   0\;\forall\;
                                                                   \gamma'\;\in\;
                                                                   \Gamma\\0 & \mathrm{otherwise} \end{array}\right.\;,
\end{align}
where $\gamma_{f_i}$ is an integral basis of flavour charges and
$f_i(\gamma)$ the flavour charges of $\gamma$. The
$\Tr[X_{\gamma_{f_i}}]$ are free quantities that are to be identified
with the flavour fugacities appearing in the index. Using the above
machinery, the 4D Schur index can be read off from the BPS quiver
\cite{Cordova:2015nma}.

In view of the similarities between the NS limit of the 5D index
discussed above and the Schur index for an $\CN = 2$ 4D theory with
the same number of vector and hypermultiplets, it is natural to wonder
whether a decomposition in terms of ``5D BPS quiver data'' also
exists. In fact, Iqbal and Vafa have used the topological string
\cite{Iqbal:2012xm} to argue that the 5D BPS-particle spectrum
reproduces the superconformal index.

We will next provide a simple but concrete realisation of this idea,
relating the NS index to the trace of the KS operator for a number of
abelian examples.  At this point we should make it clear that there
exist no nontrivial abelian fixed points in five dimensions and one
may be alarmed that the notion of the superconformal index is
ill-defined.  However, the quantity Eq.~\eqref{IRdef}, and its
subsequent NS limit, is meaningful even for non-conformal theories and
it is this definition that we will use in the upcoming
discussion.\footnote{Having said that, ``$\SU(1)$ theories" can exist
  at fixed points, since they correspond to $pq$ branewebs which can
  be collapsed to an intersection of fivebranes at a point.  For
  instance, a pure ``$\SU(1)$" theory can be engineered in the NS-D5
  intersection, and corresponds to a pure $\U(1)$ gauge theory where
  the perturbative vector multiplet is removed. The leftover instanton
  sector, behaving as a hypermultiplet, then still remains. Thus, our
  abelian computations can be understood in terms of these ``$\SU(1)$"
  theories, which often appear in quiver tails (\textit{e.g.} \cite{Bergman:2014kza, Ohmori:2015pua}).}

We have already seen that the existence of BPS instanton
particles in 5D leads to index contributions with a new global
fugacity, related to the topological charge.  It is therefore natural
to suspect that any 5D extension of the Schur--KS correspondence must
involve a BPS quiver where at least one extra node, corresponding to
the BPS instanton particle, is added. 

Unlike four dimensions, the five-dimensional central charge is real
and the BPS states are divided into CPT-conjugate pairs. The states
with the lowest possible charges (the ``partonic'' BPS states)
comprise of W-bosons and quarks, instanton solitons and
magnetically-charged BPS strings; see e.g. \cite{Aharony:1997bh}. The
existence of BPS strings makes the identification of the appropriate
five-dimensional nonabelian generalisation of the BPS-quiver
subtle.\footnote{For example, the results in \cite{Iqbal:2012xm}
  suggest that only BPS particles are important in reproducing the
  index.} However, for abelian theories with $N_f$ flavours BPS-string
states are absent and one can straightforwardly construct a 5D quiver
comprising only of an instanton-particle node and a node for each of
the $N_f$ flavours, with no arrows extending between them.

In the following section we will show that the abelian NS index can be
re-expressed to match the trace of the KS operator for the
corresponding 5D BPS quiver. We will also comment on the possible
extension to nonabelian gauge groups.

\subsection{Abelian theories}

For abelian theories the nonperturbative contribution is particularly
simple. This allows for a straightforward reinterpretation of their NS
index in terms of quiver data. The instanton partition function for
the $\U(1)$ theory with $F$ flavours and Chern--Simons (CS) level
$\kappa$ can be borrowed from \cite{Kim:2012gu}:\footnote{Compared to
  that reference, we have unrefined in the flavour fugacities for
  simplicity.}
\begin{align}\label{Nekinst}
Z_{\rm Nekrasov}^{(k)} &=\frac{(2i)^{k(F-3)}}{k!}\times\cr
& \times\int
\prod_{I=1}^k\frac{\d \phi_I}{2\pi}\frac{e^{i\kappa\phi_I} (\sin\frac{\phi_I}{2})^F\,\prod_{I\ne J}\sin\frac{\phi_I-\phi_J}{2}\prod_{I,J}\sin\frac{\phi_I-\phi_J-2i\gamma_1}{2}}{\prod_{i=1}^N\sin\frac{\phi_I-\alpha_i-i\gamma_1}{2}\sin\frac{-\phi_I+\alpha_i-i\gamma_1}{2}\,\prod_{I,J}\sin\frac{\phi_I-\phi_J-i\gamma_1-i\gamma_2}{2}\sin\frac{\phi_I-\phi_J-i\gamma_1+i\gamma_2}{2}}\;.\cr
\end{align}
Recall that, as is well-known, integrating out a massive flavour
produces a shift to the CS level by a factor of
$\Delta\kappa=\frac{{\rm sign}(m)}{2}$. As a consequence, odd $F$
requires a half-integer $\kappa$. In order to take the NS limit of the
index, we shall rewrite the above expression using the fugacities $p$
and $q$, as well as a gauge fugacity $u$:\footnote{The chemical
  potentials $\gamma_1, \gamma_2$ appearing here are not related to
  the vectors $\gamma$ of the charge lattice $\Gamma$. We hope that
  this notation, which is compatible with the literature, will not
  cause confusion.}
\begin{equation}
p=e^{-(\gamma_1 + \gamma_2)}\;,\qquad q=e^{-(\gamma_1 -
  \gamma_2)}\;,\qquad u=e^{i\alpha}\;.
\end{equation}
We will next consider specific cases by fixing the CS level and the
number of flavours.

\subsubsection*{Pure $\U(1)_{\pm 1}$ theory}

Let us consider the pure $\U(1)$ theory. The bound in
\cite{Intriligator:1997pq} requires $|\kappa|=0,1$. Setting $\kappa=1$
we find from \eqref{Nekinst}
\begin{equation}\label{noninvariant}
Z_{\rm Nekrasov}^{(1)}=\frac{1}{u}\frac{pq}{(1-p)(1-q)}\;.
\end{equation}
As discussed in \cite{Bergman:2013ala}, the instanton contributions
should be invariant under a transformation that simultaneously sends
$p\to 1/q$ and $q \to 1/p$; this is a transformation that is part of
the superconformal group, under which the perturbative single-letter
indices are invariant. However, as it stands \eqref{noninvariant} is
not invariant and this presents a problem.

Recall that this issue typically arises whenever the corresponding
brane configuration involves parallel external 5-brane legs. Indeed,
in the case of $\SU(N)_N$ theories, the brane web includes a pair of
external parallel NS5 branes. In the process of computing the
instanton contributions by decoupling the $\U(1)$ factor from the
$\U(N)_N$ theory, one finds that the naive result does not exhibit the
expected $p\to 1/q$ and $q \to 1/p$ invariance. As first argued in
\cite{Bergman:2013ala}, this noninvariance can be traced back to extra
states left over from the naive truncation, which in the brane web
description correspond to D-strings stretched between the parallel
external NS5s. These can slide off to infinity, and hence should not
be taken into account.

The discarded contribution from \cite{Bergman:2013ala} turns out to be
precisely equal to the naive $\U(1)_1$ instanton piece
\eqref{noninvariant}. As a result, going over the same brane-web
argument, we conclude that \eqref{noninvariant} corresponds to states
which should not be counted in the 5D theory. Upon removing them we
are left with $Z_{\rm Nekrasov}^{(1)}=0$, so that the full instanton
contribution in this case is simply unity. Note that had we chosen the
other sign for the CS level, $\kappa=-1$, we would have found the same
function upon taking $u\rightarrow u^{-1}$. This is tantamount to
exchanging instantons with anti-instantons, and the previous
discussion goes through unchanged.

All in all, this theory has a trivial instanton sector; the index is
purely perturbative and coincides with the Schur index of a
four-dimensional $\mathcal N = 2$ theory with the same gauge and
flavour symmetries. Since there are no BPS particles in this rank-1
theory, the corresponding 5D BPS quiver is trivial. One can therefore
simply express the answer in the general form of \eqref{KSoperator} by
writing
\begin{equation}
I_{\mathrm{KS}}=(q)^2_{\infty}\,.
\end{equation}

\subsection*{Pure $\U(1)_0$}

In four dimensions the Schur index of the pure $\U(1)$ theory at zero
CS level, $\kappa = 0$, simply reads
\begin{equation}
I^{\rm 4D}={\rm PE}\Big[-\frac{2q}{(1-q)}\Big]=\prod_{n=1}^{\infty}(1-q^{n})^2=(q)^2_{\infty}\;.
\end{equation}
In turn, the BPS quiver in 4D is trivial and therefore
\begin{align}
  \label{eq:12}
 {\rm Tr}[\mathcal{O}]=1 \;.
\end{align}
This fits the pattern of \cite{Cordova:2015nma}, since from
\eqref{KSoperator} one also recovers that $I_{KS}=(q)^2_{\infty}$. 

Let us now go to five dimensions. The exact index of the pure $\U(1)$
theory in 5D was worked out in \cite{Rodriguez-Gomez:2013dpa}. This is
\begin{equation}
\label{U1full}
I^{\mathrm{5D}}_{\U(1)_0}={\rm PE}\Big[-\frac{p+ q}{(1-p)(1-q)}+\frac{\sqrt{pq}(\mathfrak
  q+\mathfrak q^{-1})}{(1-p)(1-q)}\Big]\;.
\end{equation}
The first term is a free vectormultiplet, while the second looks like
a hypermultiplet with the gauge fugacities replaced by the instanton
fugacities, $\mathfrak q$. We can therefore use \eqref{eq:8} to infer
the corresponding NS index
\begin{align}
  \label{eq:14}
I^{\mathrm{5D, NS}}_{\U(1)_0}  = {\rm
  PE}\Big[-\frac{2q}{(1-q)}+\frac{\sqrt q (\mathfrak
  q+\mathfrak q^{-1})}{(1-q)}\Big]\;.
\end{align}
As the instanton contribution is similar to that of a hypermultiplet,
and in view of the the fact that a free hypermultiplet contributes a
flavour node to the BPS quiver \cite{Cordova:2015nma}, it is natural
to suspect that there is a 5D BPS quiver description containing one
node and yielding the correct 5D NS index.

In order to confirm this prediction, let us first pause to consider
the nonperturbative part of the index \eqref{eq:14}. Concentrating on
instantons alone, one can rewrite their contribution as
\begin{equation}\label{vortex}
{\rm PE}\Big[\frac{\sqrt q\; \mathfrak
  q}{(1-q)}\Big]=\sum_{m=0}^{\infty}\frac{(\sqrt q\; \mathfrak
  q)^m}{\prod_{k=1}^m
  (1-q^k)} = E_q(-\mathfrak q)\;,
\end{equation} 
where in the last step we used Eq.~\eqref{Eq}. As an aside, it is
interesting to observe that the above expression can be identified
with the 5D (``K-theoretic'') vortex partition function
\cite{Dimofte:2010tz}.\footnote{The second part of Eq.~\eqref{vortex}
  is to be compared with Eq.~(3.16) of \cite{Dimofte:2010tz} or its
  generalisation Eq.~(2.40).}  In fact, the NS limit of the full 5D
index can be rewritten as
\begin{equation}
I^{\mathrm{5D, NS}}_{\U(1)_0}=\prod_{n=1}^{\infty}(1-q^n)^2\prod_{n=0}^{\infty}(1-q^{n+\frac{1}{2}}\mathfrak
q)^{-1}\prod_{n=0}^{\infty}(1-q^{n+\frac{1}{2}}\mathfrak
q^{-1})^{-1}\;,
\end{equation}
which with the help of \eqref{Eq} can in turn be massaged into
\begin{equation}
\label{pureU1}
I^{\mathrm{5D, NS}}_{\U(1)_0}=(q)_{\infty}^2E_q(-\mathfrak q^{-1})E_q(-\mathfrak q)=(q)_{\infty}^2{\rm Tr}[E_q(X_{-\gamma_f})E_q(X_{\gamma_f})]\;.
\end{equation}
The above expression is consistent with it originating from a 5D
rank-1 theory with a BPS quiver consisting of a single flavour
node. The corresponding quantum torus algebra is commuting and the
formal variable $X_{\gamma_f}$ can be chosen such that
${\rm Tr}[X_{\gamma_f}]=-\mathfrak q$.

\subsection*{$\U(1)_{-\frac{1}{2}}$ with one flavour}

Our next example is a $\U(1)$ theory with one flavour at CS level
$\kappa=-\frac{1}{2}$. The 5D index reads
\begin{equation}
I^{\mathrm{5D}}_{\U(1)_{-\frac{1}{2}}}=\int \frac{\d u}{u}\,Z_{\rm pert}Z_{\rm nonpert},
\end{equation}
where $u$ is the $\U(1)$ gauge fugacity and the perturbative
contribution, after massaging \eqref{eq:8}, is given by
\begin{equation}
Z_{\rm pert}=\prod_{n=1}^{\infty}(1-q^n)^2\prod_{n=0}^{\infty}(1-q^{n+\frac{1}{2}}u)^{-1}\prod_{n=0}^{\infty}(1-q^{n+\frac{1}{2}}u^{-1})^{-1}\;.
\end{equation}

In order to find the full nonperturbative contribution, given by the
plethystic exponential of the 1-instanton term, let us begin by
looking at the latter. This is given by
\begin{align}
Z_{\mathrm{Nek}}^{(1)}=\frac{\sqrt{p q} }{(1-p)(1-q)}(1-u \sqrt{pq})\;.
\end{align}
As in the $|\kappa|=1$ case, the above
expression is not invariant under a transformation which
simultaneously sends $p\to 1/q$ and $q \to 1/p$. However, following
\cite{Bergman:2013ala} and introducing a correction factor
\begin{equation}
\Delta=\frac{q p u}{(1-p)(1-q)}\;
\end{equation}
we can write a new 1-instanton partition function in terms of
\begin{equation}
{Z'}^{(1)}_{\rm Nek}=Z_{\mathrm{Nek}}^{(1)}+\Delta=\frac{\sqrt {pq}}{(1-p)(1-q)}\;.
\end{equation}
This would suggest that the correct instanton sector contribution for
$F=1$ is the same as for the $F=0$ case
\begin{equation}
Z_{\rm nonpert}={\rm PE}\Big[\frac{\sqrt{pq}}{(1-p)(1-q)}(\mathfrak q+\mathfrak q^{-1})\Big]\;.
\end{equation}
By expanding to arbitrary order in the $q$ fugacity, it is
straightforward to check that the NS index is equivalent
to 
\begin{align}
  I^{\mathrm{5D,
  NS}}_{\U(1)_{-\frac{1}{2}}}&=(q)_{\infty}^2\sum_{k_1,k_2,r_1,r_2=0}^{\infty}
                               \frac{(-1)^{k_1+k_2+r_1+r_2}q^{\frac{k_1+k_2+r_1+r_2}{2}}(-\mathfrak  q)^{r_2-r_1}\delta_{k_1,k_2}}{(q)_{k_1}(q)_{k_2}(q)_{r_1}(q)_{r_2}}\cr
  & =  (q)_{\infty}^2 \Tr[E_q(X_{-\gamma_f}) 
    E_q(X_{-\gamma}) E_q(X_{\gamma_f}) E_q(X_\gamma)]\;.
\end{align}
In complete analogy with our previous discussion, the interpretation
of this result in the language of \cite{Cordova:2015nma} would be that
the instanton provides a flavour charge $\gamma_f$, in addition
to the charge lattice vector for the hypermultiplet, $\gamma$. This is
consistent with having a 5D BPS quiver involving two nodes and no
adjoining arrows.

\subsection*{Maximally SUSY theory}

Consider adding to the U(1) vectormultiplet a hypermultiplet in the
adjoint representation. This is the content of the maximally
supersymmetric theory.\footnote{Although this theory has
  $\mathcal N=2$ supersymmetry, we can still study it using 5D
  $\mathcal N=1$ tools.} One might naively think that the adjoint
hypermultiplet decouples and as a result that the instanton
contribution is simply that of the pure $\U(1)$ theory. This is
however not the case, as the noncommutative deformation regulating the
Nekrasov partition function couples zero modes of the $\U(1)$ adjoint
hypermultiplet to the instantons. In fact, it turns out
\cite{Kim:2011mv} that the instanton contribution is
\begin{align}
\label{eq:inst}
  Z_{\rm inst}={\rm
  PE}\Big[\sum_{k = 1}^\infty \mathfrak{q}^k\,z_{\rm sp}\Big]
  \qquad\textrm{with}\qquad z_{\rm sp}=-\frac{p+q}{(1-p)\,(1-q)}+2\frac{\sqrt{pq}}{(1-p)\,(1-q)}\;.
\end{align}
As stressed in \cite{Kim:2011mv}, $z_{sp}$ is equal to the
contribution of a 6D tensor multiplet. This constitutes a nontrivial
check for the conjectured UV self-completion of the maximally SUSY 5D
theory into the $(2,0)$ theory
\cite{Douglas:2010iu,Lambert:2010iw}. Note that the expression for
$z_{\rm sp}$ above is exactly that of an abelian vector plus an adjoint
hypermultiplet. The latter is the full perturbative contribution of the 5D
maximally SUSY theory, i.e. 
\begin{align}
  \label{eq:13}
   Z_{\rm pert} = {\rm PE}[ z_{\rm sp}]\;.
\end{align}
Moreover, in the NS limit on can re-express
\begin{align}
  \label{eq:pert}
  {\rm PE}[\mathfrak{q}^k\,f_H]=\Big(\prod_{m=0}^{\infty} (1-\mathfrak{q}^k q^{m+\frac{1}{2}})^{-1} \Big)^2 =(E_q(-\mathfrak{q}^k))^2\;,
\end{align}
while
\begin{align}
{\rm PE}[\mathfrak{q}^k\,f_V]=\prod_{m=0}^{\infty} (1-\mathfrak{q}^k\,q\, q^{m})^2=(\mathfrak{q}^k\,q;\,q)^2\;,
\end{align}
where $(a;\,b)$ stands for the q-Pochhammer symbol.\footnote{The
  $(a;\,b)$ q-Pochhammer symbol is defined as $(a;\,b):=
  \prod_{j=0}^{\infty}(1-ab^j)$.} The full index is given by
\begin{align}\label{N2index}
I^{\rm MaxSUSY}_{\U(1)}= Z_{\rm pert}Z_{\rm inst}\overline Z_{\rm
  inst}\;,
\end{align}
where the overline implies an inversion of the instanton
fugacity. This prescription---which we stress is just the direct
implementation of the results of \cite{Kim:2012gu}, and strongly
supported by non-trivial checks, including the emergence of the
enhanced flavour symmetries in the case of $E_{N_f+1}$
theories---amounts to writing
\begin{align}
  \label{eq:Spole}
  \overline Z_{\rm inst}= {\rm PE}\Big[ \sum_{k=1}^{\infty} \mathfrak q^{-k}z_{\rm sp}\Big]
\end{align}
and Eq.~\eqref{N2index} can be nicely repackaged into\footnote{Note
  that by taking into account Eqs.~\eqref{eq:inst},~\eqref{eq:13}
  and~\eqref{eq:Spole} and naively resumming the instanton expansions,
  it looks like the total partition function is ${\rm
    PE}[0]=1$.
  However this conclusion is incorrect, since for this to happen each
  series is implicitly resummed in a different regime, while here
  $| \mathfrak q| = 1$.}
%
\begin{align}
I^{\rm MaxSUSY}_{\U(1)}=\prod_{k=-{\infty}}^{\infty}(\mathfrak{q}^{k}\,q;\,q)^2\,E_q(-\mathfrak{q}^{-k}) E_q(-\mathfrak{q}^k)\;.
\end{align}
This expression does not have a strict 5D BPS quiver
interpretation. However, its form is rather suggestive: the collection
of instantons corresponds to BPS states at threshold associated with
the Kaluza--Klein modes that uplift the theory to 6D
\cite{Kim:2011mv}. As such, one may expect that these would provide an
infinite tower of flavour nodes, each parametrised by integer
multiples of a fundamental charge, $\mathfrak q^n$, which is what we
seem to find. However, the q-Pochhammer symbol, expected to arise from
the vectormultiplet contribution, also depends on $\mathfrak q^n$. It
is tempting to speculate that this is due to the flavour fugacity
combinations $\mathfrak{q}^n$ being remnants of a 6D Lorentz fugacity.

\subsection{Towards nonabelian theories}

It is natural to ask whether there exists a nonabelian extension of
the correspondence between the NS index of a 5D SCFT and the trace of
the KS operator for an associated BPS quiver, but we have thus far not
been successful in constructing any such examples. Having a
closed-form expression for the nonperturbative part of the nonabelian
NS index---the perturbative part reduces trivially to the 4D Schur
index---would be helpful in pursuing this direction. Although the NS
limit of the abelian K-theoretic Nekrasov partition function coincides
with the K-theoretic vortex partition function---{\it c.f.}  under
Eq.~\eqref{vortex}---explicitly applying our prescription
\eqref{NSlimit} to nonabelian gauge groups quickly produces an answer
which disagrees with the $\mathfrak q$-expansion of any K-theoretic
vortex partition function.

However, there may be another way forward using dualities. The
instanton partition function---the 4D limit of the nonabelian 5D
Nekrasov partition function\footnote{This is known as the
  ``homological limit'' (see \textit{e.g.} \cite{Dimofte:2010tz} and
  references therein) and in notation where one has made explicit the
  dependence of the fugacities on the Euclideanised time radius,
  $p = e^{- \beta \epsilon_1}$, $q = e^{-\beta \epsilon_2}$,
  corresponds to taking $\beta\to 0$. In this limit, the full
  ``K-theoretic'' version of the Nekrasov partition function we have
  been using thus far reduces to the 4D instanton partition
  function.}---has a well-defined NS limit, originally discussed in
\cite{Nekrasov:2009rc,Nekrasov:2012xe,Nekrasov:2013xda}, of which our
prescription \eqref{NSlimit} is a natural generalisation. As can be
seen by expanding in the instanton fugacity, and simultaneously for
small but nonzero $\epsilon_1, \epsilon_2$, Eq.~\eqref{eq:10} becomes
\begin{align}
  \label{eq:11}
Z_{\rm Nek}^{\rm NS} & \to  1 + \mathfrak q  \lim_{\epsilon_2 \to 0 } \epsilon_2 Z_{\mathrm{inst}}^{(1)}
  + \lim_{\epsilon_2 \to 0 } \mathfrak q^2\Big(\frac{\epsilon_2 (\epsilon_2-1)}{2}(Z_{\mathrm{inst}}^{(1)})^2
   + \epsilon_2 Z_{\mathrm{inst}}^{(2)}\Big) + O(\mathfrak q^3)\cr
& =  \exp \Big[\lim_{\epsilon_2 \to 0}
  \epsilon_2 
  \log Z_{\mathrm{inst}}(\mathfrak q;\epsilon_1,\epsilon_2)\Big] \;,
\end{align}
precisely the expression appearing in \cite{Nekrasov:2009rc}. In that
reference, the resultant partition function was identified as the
nonperturbative contribution to the twisted superpotential for some
associated two-dimensional theory. Subsequently, the authors of
\cite{Dorey:2011pa,Chen:2011sj} also linked the full 2D twisted
superpotential---the NS limits of the full perturbative plus
nonperturbative partition functions of the 4D theory---with the
twisted superpotential for a different, dual 2D theory. Interestingly,
the latter theory can in certain cases---\textit{e.g.} the abelian
example---be interpreted as the worldvolume description for a 2D
defect in the Higgs branch of the original 4D theory. The partition
function for these defects is the well-known vortex partition
function, which has a natural K-theoretic lift up to 5D.\footnote{The
  details of this K-theoretic vortex partition function depend on how
  the embedding of the defect breaks the gauge group, {\it i.e.} on
  the choice of Levi subgroup; see \textit{e.g.}
  \cite{Dimofte:2010tz} and references therein for details. See also
  \cite{Kim:2016qqs} for new results on the calculation of 5D
  instanton partition functions in the presence of defects.} It would
be interesting to closely study similar
5D$\to$4D$\to$2D$\to$2D$\to$4D$\to$5D chains for more complicated
theories. This in turn could lead to identifying closed-form
expressions for the NS limits of nonabelian instanton contributions
and shed light on how to proceed with the nonabelian extension of the
NS-KS correspondence presented in this section.

Another closely-related task would be to investigate whether the 5D NS
index we have defined admits an alternative (and possibly simpler)
description associated with some lower-dimensional structure, along
the lines of \cite{Beem:2013sza,Beem:2014kka}. In this respect, the
similarity of our prescription to the large orbifold limit of
\cite{Mekareeya:2013ija} may hint towards such a connection. We will
leave the answers to these questions as open problems for future
research.

\ack{ \bigskip We would like to thank M.~Buican and J.~Hayling for
  helpful discussions and comments. A.P. and D.R.G. would like to
  acknowledge the CRST at Queen Mary University of London for
  hospitality during various stages of this work and the associated
  support of the COST Action MP1210 STSM.  C.P. is supported by the
  Royal Society through a University Research Fellowship. A.P. and
  D.R.G. are partly supported by the Spanish Government grant
  MINECO-13-FPA2012-35043-C02-02. In addition, they acknowledge
  financial support from the Ramon y Cajal grant RYC-2011-07593 as
  well as the EU CIG grant UE-14-GT5LD2013-618459. The work of A.P is
  funded by the Asturian Government SEVERO OCHOA grant BP14-003.}

\newpage



\bibliography{NSSchur}

\end{document}